# Simulation Guided Molecular Design of Hydrofluoroether Solvent for High Energy Batteries

Zhou Yu[1,2,+], Zhangxing Shi[1,3,+], Sambasiva R. Bheemireddy[1,3], Ethan Kamphause[1,2], Xingyi Lyu[4], Mohammad Afsar Uddin[1,5,10], Zhiguang Li[1,3,7], Zhenzhen Yang[1,3], Tao Li[1,4,6], Jeffrey S. Moore[1,5,8,9], Lu Zhang[1,3*], Lei Cheng[1,2*]

[1]Joint Center for Energy Storage Research, Argonne National Laboratory, Lemont, IL 60439, USA.
[2]Materials Science Division, Argonne National Laboratory, Lemont, IL 60439, USA
[3]Chemical Sciences and Engineering Division, Argonne National Laboratory, Lemont, IL 60439, USA
[4]Department of Chemistry and Biochemistry, Northern Illinois University, DeKalb, IL 60439, USA
[5]Department of Chemistry, University of Illinois at Urbana–Champaign, Urbana, IL 61801, USA
[6]X-ray Science Division, Argonne National Laboratory, Lemont, IL 60439, USA
[7]Indiana University-Purdue University Indianapolis, 723 West Michigan Street, Indianapolis, IN, 46202, USA
[8]Beckman Institute for Advanced Science and Technology, University of Illinois at Urbana–Champaign, Urbana, IL 61801, USA
[9]Department of Materials Science and Engineering, University of Illinois at Urbana–Champaign, Urbana, IL 61801, USA
[10]Instituto de Ciencia de Materiales de Madrid, Madrid 20849, Spain

[+]Zhou Yu and Zhangxing Shi contributed equally to this work.
*Corresponding Authors: Lu Zhang luzhang@anl.gov and Lei Cheng leicheng@anl.gov

**TOC**

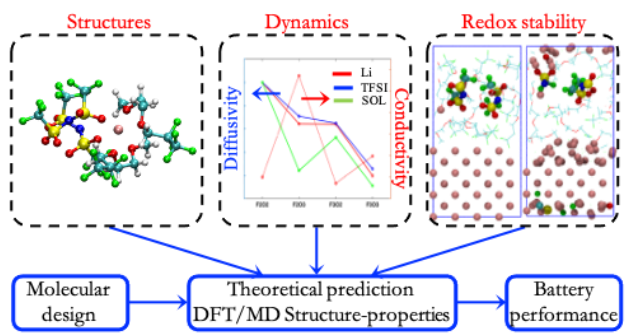

Abstract: Electrolyte design is critical for enabling next-generation batteries with higher energy densities. Hydrofluoroether (HFE) solvents have drawn a lot of attention as the electrolytes based on HFEs showed great promise to deliver highly desired properties, including high oxidative stability, ionic conductivity, as well as enhanced lithium metal compatibility. However, the structure-dynamics-properties relationships and design principles for high-performance HFE solvents are still poorly understood. Herein, we proposed four novel asymmetric HFE designs by systematically varying polyether and fluorocarbon structural building blocks. By leveraging molecular dynamics (MD) modeling to analyze the solvation structures and predict the properties of the corresponding 1 M lithium bis(fluorosulfonyl)imide (LiTFSI) solutions, we downselected the most promising candidate based on high conductivity, solvation species distribution, and oxidative stability for extensive electrochemical characterizations. The formulated electrolyte demonstrated properties consistent with the predictions from the simulations and showed much-improved capacity retention as well as Coulombic efficiency compared to the baseline electrolytes when cycled in lithium metal cells. This work exemplifies the construction of candidate electrolytes from building block functional moieties to engineer fundamental solvation structures for desired electrolyte properties and guide the discovery and rational design of new solvent materials.

Increasing the energy density of batteries is highly desired as the world market of electric vehicles (EVs) proliferates. While lithium-ion batteries (LIBs) are still the most widely adopted energy storage solution for the application, the current chemistry is approaching its theoretical limits. Lithium metal anode has a theoretical capacity of 3860 mA h g$^{-1}$ and the lowest reduction potential (i.e., -3.04 V vs. SHE), making lithium metal battery (LMB) an attractive option with a doubled energy density compared to LIBs.[1] However, LMBs still have formidable challenges in battery stability and safety, both of which are closely related to lithium anode issues, including dendrite growth, dead lithium, etc.[2] While the widely used carbonate-based electrolytes do not mitigate those issues, electrolyte engineering is a critical approach for LMBs as certain formulations form a full passivation film on the Li surface, leading to enhanced solid electrolyte interphase (SEI) stability.

To this end, a broad range of parameters are available to tune electrolyte behavior or properties, including the composition of salt and solvent, concentration, temperature, solvent structure design, etc. [3-11] For example, salt-concentrated electrolytes have been shown to improve the performance of LMBs [6, 12] as the high salt concentrations suppress the growth of dendrites through stabilizing the ionic concentration profiles[13], and facilitate the formation of the stable anion-derived SEI [14, 15]. The realized benefits of such electrolytes were attributed mainly to the unique solvation structure originating from the high concentration. The localized high-concentration electrolytes (LHCEs) have shown similar solvation structure features and performance enhancement without the high concentration of salts, which impairs conductivity and adds cost. However, the oxidative stability of the LHCEs is still limited.[8, 9]

Novel solvents have also been heavily investigated, including variations of carbonates, nitriles, ethers, and sulfones.[11] Among them, ether-based electrolytes show the most reductive resistance due to the lack of reduction-sensitive double bonds. However, for the same reason, they also suffer limited oxidative stability.[16] Introducing fluorine to ether backbones is an effective approach to increasing oxidative stability. Recently, several electrolytes based on hydrofluoroether (HFE) solvents have shown promising properties, including high oxidative stability, fast ionic conductivity, and enhanced lithium metal compatibility.[7, 8, 17, 18] For instance, fluorinated orthoformate, when used as a co-solvent in LMBs facilitates the formation of highly homogeneous monolithic SEI, suppressing Li dendrite growth.[17] Another synthesized fluorinated ether, DEG-FtriEG, demonstrates both high ionic conductivity (2.7×10$^{-4}$ S/cm) and high oxidative stability (5.6 V).[8] A recently reported HFE solvent, fluorinated 1,4-dimethoxylbutane, exhibits surprisingly excellent compatibility with Li metal and outstanding LMB performance when used in a single solvent electrolyte of 1 M Lithium bis(fluorosulfonyl)imide (LiFSI).[7] Despite the exciting progress and attempts to reveal an in-depth understanding of the solvation impact, systematic investigation of the effects of building block groups of fluorocarbons and polyethers on the solvation structure, ion dynamics, and battery performance is absent, and the solvation-property relationships of the electrolyte at a molecular level are elusive.

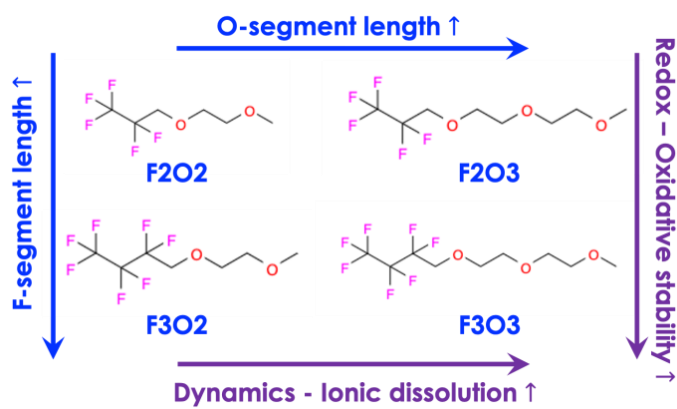

Scheme 1. Structures of novel fluorinated ethers with varied polyether and fluorocarbon segments that we investigated in this work.

In this study, we adopted an asymmetric design and proposed a new family of HFE solvents by varying the lengths of fluorocarbons and polyethers segments, as shown in Scheme 1. Unlike other reported HFE structures, the polyethers were introduced into fluorocarbons only on one side or asymmetrically, intentionally creating intramolecular dipole moments that could benefit the solvating capability of Li ions. The notation of those structures (FxOy with x,y=2,3) was defined using the numbers of fluorinated carbon and oxygen atoms incorporated. To accelerate the discovery process, density functional theory (DFT) calculations, *ab initio*, and classical molecular dynamics (AIMD/CMD) simulations were used to establish the *in silico* screening protocols toward key properties of electrolytes composing of 1 M LiTFSI in those HFE solvents. Specifically, solvation structure, transport properties, redox stability of the electrolytes, and their reactivity with Li anode were investigated. The experimental characterizations were followed to validate the results. Based on this approach, the most promising candidate, F2O3, was selected for extensive battery performance characterization, showing improved cycling performance. The LMB cell using 1 M LiTFSI in F2O3 as the electrolyte with a metallic Li anode and an NMC622 cathode showed high average Coulombic efficiency and excellent capacity retention for 100 cycles at a C/3 rate. This work highlights the adoption of *in silico* design protocols to accelerate the discovery loop of HFE solvents for LMBs through the understanding of the solvation-property relationship and further tailoring the chemical design to tune solvation structures.

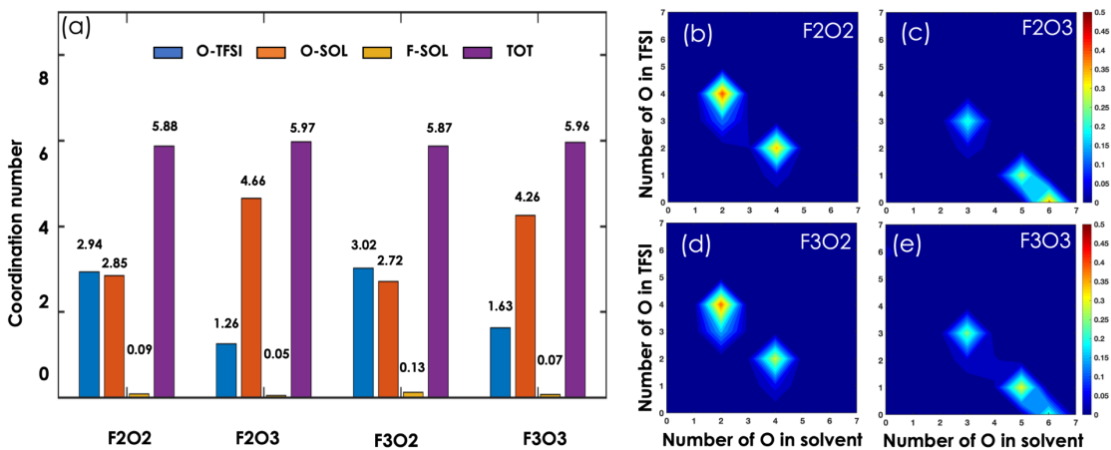

Figure 1. Solvation structure analysis of 1 M LiTFSI in F2O2, F2O3, F3O2, and F3O3 solvents. (a) Coordination number (CN) of Li ions with Os of TFSI ions (O-TFSI), Os of solvent molecules (O-SOL), and Fs of solvent molecules (F-SOL) in different electrolytes. The total CN is represented by TOT. (b-e) Time-averaged appearance frequency of the CNs of Li ions by Os in TFSI ions and solvent molecules in different electrolytes. The color bar represents the occurrence frequency.

*Solvation Structure*

The solvation structure of electrolytes is closely related to their properties. Hence, we first characterize the solvation structures of 1 M LiTFSI in HFE solvents via MD simulations (Simulation details are in supporting information). From the radial distribution function (RDF) analysis (Figure S2), we see that Li ions are coordinated mainly by O atoms in TFSI ions (O-TFSI) or solvent molecules (O-SOL), and the contribution from F atoms in solvent molecules (F-SOL) within the first solvation shell is very small (<3%). From averaged coordination number analysis shown in Figure 1a, as the polyether segment increases from F2O2 electrolyte to F2O3 electrolyte or from F3O2 electrolyte to F3O3 electrolyte, the coordination numbers (CNs) of O-TFSI decrease, and the CNs of O-SOL increase, while the total CNs (TOT) are similar for all solvents.

Time-averaged appearance frequency analysis of the CNs represented using 2D density maps (Figure 1b-e) reveals more details on dominant solvation compositions. The figures show that the most common solvation structures in all four electrolytes have both O-TFSI and O-SOL coordination, but the exact numbers vary as the solvents change. In F2O2 and F3O2 electrolytes, the preferred Li ion coordination environments (identified by the high occurrence frequency areas, see color bar) consist of either 2 O-SOLs and 4 O-TFSIs or 4 O-SOLs and 2 O-TFSIs, with the former being more favorable. In F2O3 and F3O3 systems, the majority of Li ions are coordinated with 5 or 6 O-SOLs, while a small percentage are coordinated with a combination of 3 O-SOLs and 3 O-TFSIs. By further examining the coordination species (Figure S3), we observe that the coordinated Os in the solvation shell are from 1~2 solvent molecules and 2~3 TFSI ions in the F2O2 and F3O2 electrolytes. As the polyether chains increase as in F2O3 and F3O3 electrolytes, the Li ions are more likely to be coordinated by either just two solvent molecules or two solvent molecules plus 1 TFSI ion. This observation is consistent with the more substantial chelating effect of the increased polyether length. The three most prominent and representative solvation structures in each electrolyte in MD simulations are shown in Figure S4.

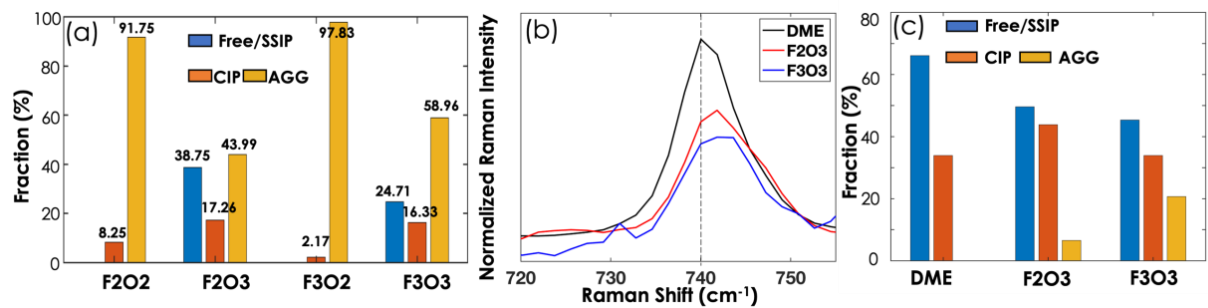

Figure 2. (a) Fraction of ions in free ion/solvent-separated ion pair (Free/SSIP), contact ion pair (CIP), and aggregate (AGG) solvation structure. (b) Raman spectral comparison of the TFSI breathing model region for solutions of 1 M LiTFSI in F2O3, F3O3 and DME solvents. The frequency of the free/SSIP TFSI model is indicated by a dashed line to guide the eyes. (c) Calculated percentage of the integrated Raman signal intensity in each electrolyte.

The speciation of ions is further categorized based on their local solvation structure as free ion/solvent-separated ion pair (Free/SSIP), contact ion pair (CIP), and aggregate (AGG), in which these ions are coordinated by zero, one, or more than one counterions, respectively.[19, 20] From Figure 2a, we can see more than 90% of Li ions form AGG in F2O2 and F3O2. As the polyether chain increase in length, the AGG formation is suppressed, and the fraction of Free/SSIP and CIP increases. On the other hand, the increase of fluorinated moieties (F3Oy vs. F2Oy) seems to correlate with an increased AGG fraction. The speciations of 1 M LITFSI in F2O3 and F3O3 were also studied by Raman. The synthesis details of F2O3 and F3O3 were described in the supporting information. Previous literature studies have used Raman measurements of the TFSI breathing mode at peak positions around 740-750 cm$^{-1}$ to characterize and quantify speciations in various electrolyte systems.[21] We fitted the peaks from our experiments (Figure 2b) following the previous literature, assigning the free ion peak at 740 cm$^{-1}$, contact ion pair (CIP) peak at 745 cm$^{-1}$, and aggregate (AGG) peak at 750 cm$^{-1}$ [14, 22, 23] (details on spectral deconvolutions shown in Figure S5). The resulting percentages from the integrated Raman intensities are shown in Figure 2c. The observation of more AGG in F3O3 electrolyte than in F2O3 electrolyte is qualitatively consistent with MD simulation, validating the effect of molecular design predicted by simulations.

Besides the well-studied local solvation structures, detailed analysis of the aggregates is necessary as the nanometric aggregate (n-AGG) has recently been observed in electrolytes for various battery chemistries and these long-range structures inevitably influence ion transport, redox characteristics, and mechanics.[20] The aggregate structure analysis (Figure S7) shows that small ionic aggregates with several ions are the most common, while n-AGG with several tens of ions also exists in the electrolytes. The increase of polyether segments of those solvents correlates with the reduced sizes of n-AGG. Aggregate structures are more commonly observed in electrolytes with high salt concentrations where the systems are close to the solubility or precipitation limit. In the HFE solvents studied here, these limits are low because of the weaker solvation ability of HFE and strong solvent-solvent interactions resulting from the fluorous effect. For example, a majority (> 85%) of F atoms in solvent molecules are in close contact with at least another F atom of a different solvent molecule, as shown in Table S3 and Figure S8. Such solvent-solvent interactions limit the flexibility of solvents to coordinate freely with salt, contributing to the relatively low solubility of salt in HFE.

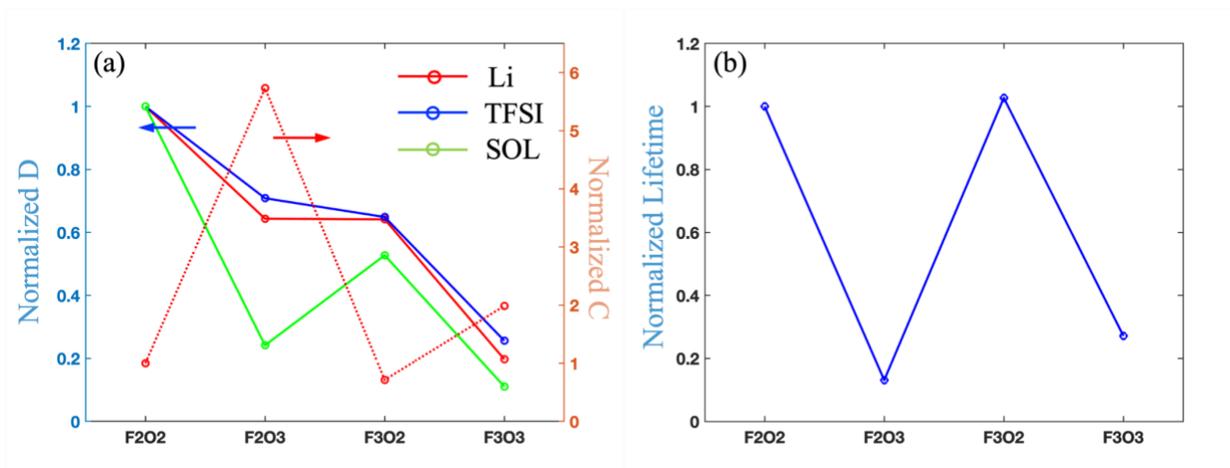

Figure 3. (a) Self-diffusivity (D) of Li ion, TFSI ion, and solvent molecule and ionic conductivity (C); (b) Lifetime of Li TFSI ion pair. All data is normalized by the F2O2 system and the exact data is summarized in Table S4.

*Dynamics Properties*

The second step of this *in silico* design is to estimate the dynamic properties of electrolytes, which are vital to battery performance. From the macroscopic standpoint, we calculated the self-diffusivity of cation, anion, and solvent and the ionic conductivity of the electrolyte. The results are reported in Table S4, and the normalized results are shown in Figure 3a. The calculation methods were introduced elsewhere.[24] We observe that as the size of solvent molecules increases, the self-diffusivities of ions and solvents decrease, which is consistent with the Stokes-Einstein relation that a larger solvent molecule slows the electrolyte dynamics. Increasing the polyether length correlates positively to the calculated ionic conductivity, as F2O3 and F3O3 electrolytes have higher conductivity compared to those of F2O2 and F3O2 electrolytes. F2O3 electrolyte has a significantly higher conductivity, which is nearly three times the second-highest value of F3O3 electrolyte. We note that the F2O2 electrolyte has the highest ion and solvent diffusivities but relatively low ionic conductivity. The strong correlation among species in the electrolytes leads to this highly nonideal behavior. From the microscopic standpoint, we quantify the dynamics of electrolytes by the lifetime of Li-TFSI ion pairs and the transport mechanism of ions. As shown in Figure 3b, the lifetime of Li-TFSI ion pairs in different electrolytes is calculated by fitting the Li-TFSI association correlation function shown in Figure S9 (See the SI for details). The longer polyethers reduce the lifetime of the Li-TFSI ion pair, which correlates negatively with the calculated ionic conductivity. Thus, the cation-anion association in CIP or AGG is responsible for the decreased ionic conductivity of the electrolytes.

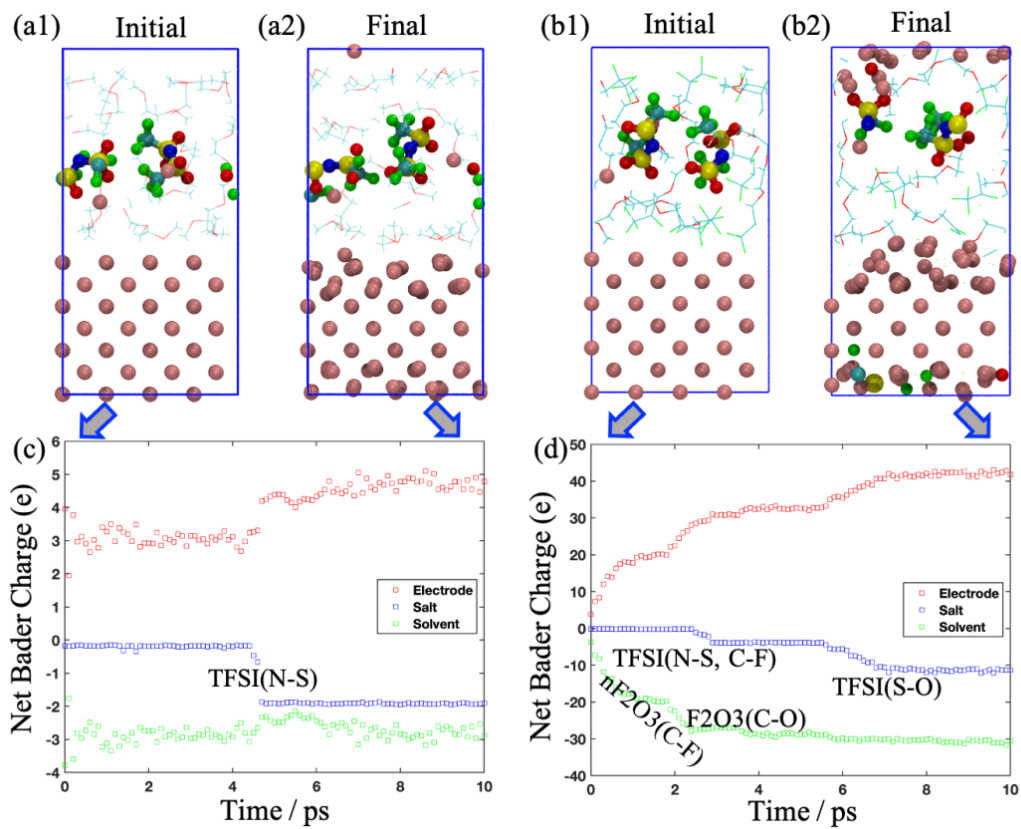

Figure 4. The initial and final snapshots extracted from AIMD simulations with (a) DME as solvent or (b) F2O3 as solvent. LiTFSI locates in the middle of the electrolyte in the initial configuration. Pink, blue, yellow, red, cyan, and green balls denote Li, N, S, O, C, and F, respectively. The solvent molecules are depicted as wireframe. Net Bader charges of components in systems with (c) DME as solvent or (d) F2O3 as solvent.

*Redox Properties and SEI formation*

Another important step is to estimate the redox characteristics of the electrolytes and how they affect SEI formation. The redox properties were firstly determined through the analysis of HOMO and LUMO energies from DFT calculations shown in Table S5. It is clear that the fluorinated ethers have lower HOMO and LUMO levels as compared to a non-fluorinated ether solvent dimethoxyethane (DME). Lower HOMO corresponds to improved oxidative stability, which is a known effect of fluorination. Lower LUMO indicates that the molecule is more readily reduced, presenting an opportunity to tune the solid electrolyte interphase (SEI) layer towards prolonged cycle life[25] and reduced interfacial resistance[26] through solvent formulation. The higher fraction of CIP and AGG in the fluorinated electrolytes (see Figure 2) compared to the DME system (see Figure S10) also suggest lower LUMO energy levels of the TFSI ion affected by the coordinated

Li ion[6] and a greater likelihood of reduction, which may promote an anion-derived SEI layer that is known to benefit SEI passivation.[25, 26]

The interfacial reactions that initiate the formation of the SEI layer are studied via AIMD simulations of the electrolytes in contact with a Li metal anode. We perform simulations on DME, F2O3, and F3O3 electrolytes for comparison. The initial and final structures and the Bader charge analysis results of these simulations are reported in Figure 4. In the 1M LiTFSI in DME electrolyte, DME remains unreactive within the 10 ps of simulation time (Figures 4a and 4c), which is consistent with the recent literature study.[27] Despite not being in direct contact with the Li surface, the LiTFSI salt reacts within the short simulation time. The reduction and decomposition of TFSI ion via N-S is observed at ~4.50 ps, and 2 electrons transfer from the metallic Li electrode to the salt during the cleavage of N-S (Figure 4c). As shown in Figure 4b, the decomposition of F2O3 via C-F bond breaking occurs within 0.5 ps of simulation time, and the cleavage of the C-O bond is observed after 5 ps. The decomposition of TFSI ion via N-S, C-F, and S-O also occurs fairly early before 3ps. The prominent charge transfer from the electrode to solvent and salt is shown in Figure 4d. We note that the cleavage time of a certain type of bond may slightly change with different initial configurations. For example, in simulations where the salt species are in contact with the Li surface, salt and F2O3 decomposition also occur within 10 ps of simulation but at different times (see Figure S11). In the F3O3 system shown in Figure S12, the C-F bond of the HFE breaks within the initial 100 fs, and C-O bond breakage is observed within the first 0.5 ps for some species. The reduction is not limited to a single species but occurs for multiple HFEs throughout the trajectory. The overall amount of charge transferred to the solvent from the Li electrode is similar between the F2O3 and the F3O3 case but occurs on a faster time scale for F3O3. Despite the similarities in charge transfer and reactions with the HFE, the TFSI molecules do not undergo a reduction in the F3O3 case. The overall charge of TFSI molecules remains close to neutral, and the bonds/structure of the molecule intact. This indicates that the reduction of F3O3 is more favorable than the TFSI and the F2O3. Nevertheless, we highlight that the decomposition of newly synthesized fluorinated ether is more facile than DME due to the strong electron-withdrawing effect of the F functional group[28], likely leading to fluorine-rich SEI layers that can be beneficial to stabilizing Li metal anodes.

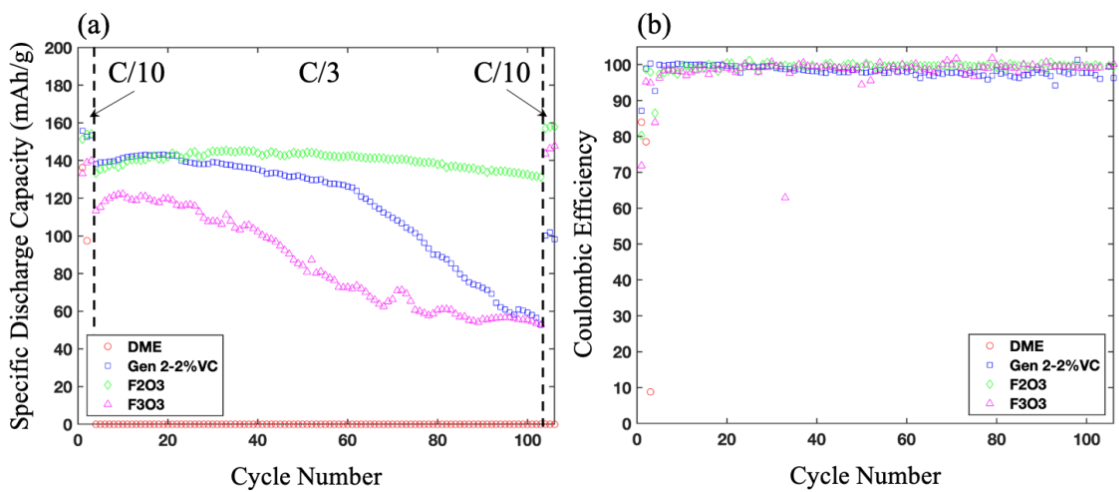

Figure 5. (a) Discharge capacity and (b) Coulombic efficiency profiles of as a function of cycle number for cells containing 1 M LiTFSI DME/F2O3/F3O3 electrolytes and 1 M LiPF$_6$ in EC/EMC (3/7) and 2% vinylene carbonate.

*Battery performance validation*

Based on the simulation results of solvation structures, dynamic properties, redox characteristics, and associated possible SEI involvements, we identify the F2O3 electrolyte as the most promising candidate as it possesses balanced properties including high conductivity, high oxidative stability as well as feasible SEI reactivity. As expected, both F2O3 and F3O3 electrolytes show significantly improved oxidative stability compared to the DME electrolyte. In the linear sweep voltammetry (LSV) tests (Figure S13), both electrolytes postpone the decomposing currents until above 5 V vs. Li/Li$^+$, while the DME electrolyte exhibits increased decomposition at around 4.3 V vs. Li/Li$^+$. While F2O3 and F3O3 electrolytes start to decompose at the same potential, the F3O3 electrolyte has a much lower decomposition current, indicating its better oxidative stability. Interestingly, despite the structural similarity, the F2O3 electrolyte exhibits a significantly higher ionic conductivity of 1.81 mS/cm from the electrochemical impedance spectroscopy (EIS) measurements (Figure S14) compared to that of the F3O3 electrolyte at 0.76 mS/cm. These results agree well with the predicted values (See Tables S4 and S5).

To further demonstrate the impact on cycling performance, those electrolytes undergo extensive electrochemical evaluations in Li metal cells using a LiNi$_{0.6}$Mn$_{0.2}$Co$_{0.2}$O$_2$ (NMC622) cathode. Figure S15 summarizes the capacity retention and Coulombic efficiency (CE) profiles of cells using an F2O3-based electrolyte (1 M LiTFSI in F2O3), an F3O3-based electrolyte (1 M LiTFSI in F3O3), a standard carbonate-based electrolyte (1.2 M LiPF$_6$ in EC/EMC= 3/7, plus 2 wt % vinylene carbonates, Gen 2-2%VC), as well as a DME electrolyte (1 M LiTFSI in DME). The 2032-type stainless steel coin cells used for evaluation consist of a lithium metal anode, a microporous polypropylene separator (Celgard 2325), an NMC622 cathode, and 25 μL of electrolyte. Details regarding cell assembly and cycling parameters can be found in the supporting information. The cells were subjected to three formation cycles at a C/10 rate followed by 100 cycles at a C/3 rate between 3.0 to 4.2 V. The DME cell shows very poor cycling as the capacity drops to nearly 0 within 4 cycles, possibly due to the cathode-electrolyte incompatibility. The Gen 2-2%VC cell shows better capacity retention, but only 38% capacity remains after 100 cycles. While the capacity retention of the F3O3 cell is even worse, the F2O3 cell shows a much-improved capacity retention with no obvious decay over 100 cycles. As for the CE profiles, DME cell drops from 80% to 0 within 4 cycles, while the Gen 2-2%VC, F2O3, and F3O3 cells show comparable values. F2O3 cell outperforms with a more stable profile and an average CE of 99.44%. Figure S15 compiles the voltage-capacity profiles of those cells over 100 cycles. The DME cell shows decent charge-discharge curves for the first cycle but quickly fails on the second charge, suggesting severe decomposing reactions of DME when the cathode potential is raised. Both Gen2-2%VC and F3O3 cells demonstrate increased overpotential as capacity drops over cycling. Only F2O3 cell exhibits low and stable overpotential over 100 cycles. The results suggest that besides the high conductivity, a stable SEI may form from the F2O3 electrolyte, evidenced by the limited overpotential increase and stable capacity retention.

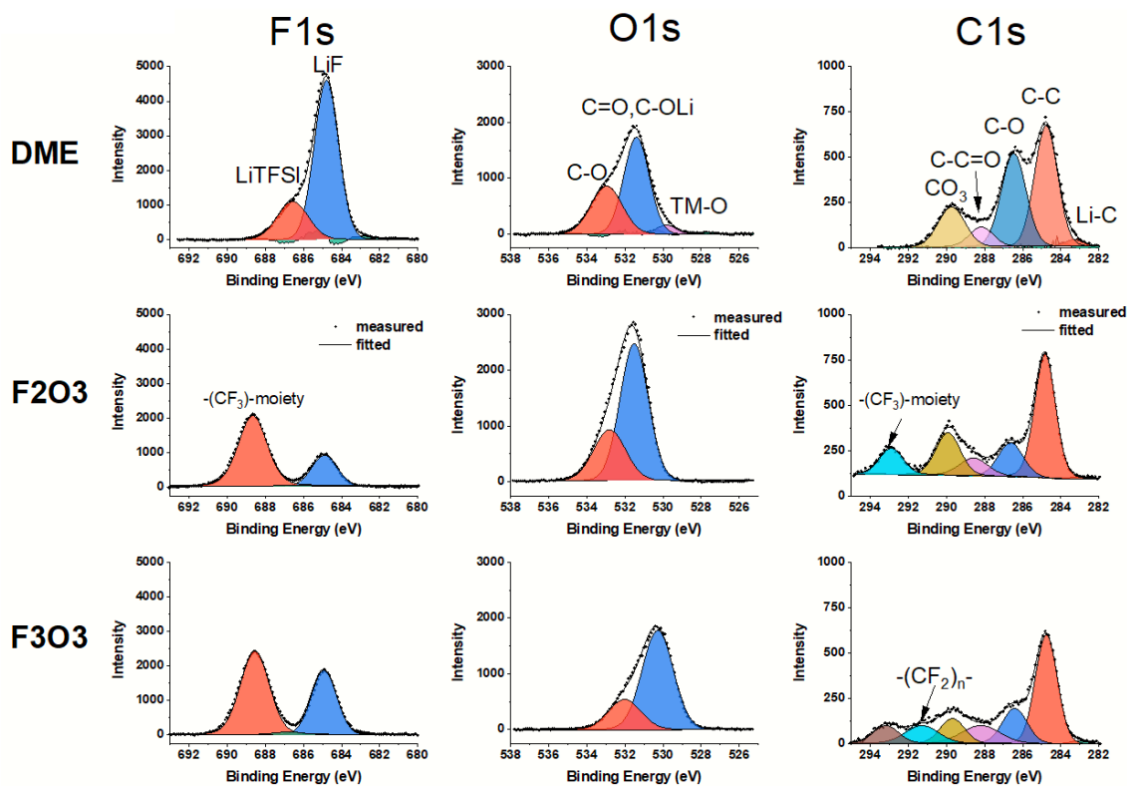

Figure 6. F 1s, O 1s, and C 1s XPS spectra of the Li metal electrodes after 100 cycles.

To understand the SEI evolution, the cycled Li metal electrodes are investigated using X-ray photoelectron spectroscopy (XPS). Figure 6 shows the F 1s, O 1s, and C 1s spectra of the cycled Li metal electrodes after 100 cycles. In the F 1s spectra, the electrode cycled with DME electrolyte exhibits an overwhelming LiF peak along with a LiTFSI peak. While inorganic LiF is known to strengthen the SEI layer, given DME mainly suffers severe decomposing on the cathode surface during high voltage cycling, the SEI on cycled Li metal surface may not provide enough justifications regarding the cell performance. On the other hand, both F2O3 and F3O3 electrodes show a LiF peak as well as a –(CF$_3$)-moiety peak, suggesting both solvents promote the CF composition during the SEI formation process. A similar observation is also found in the C 1s spectra. F2O3 and F3O3 electrodes show additional C-F peaks than the DME electrode. Interestingly, F2O3 exhibits only a –(CF$_3$)-moiety peak, and F3O3 shows both a –(CF$_3$)-peak and a –(CF$_2$)-peak. Considering –(CF$_2$)- should most likely come from HFE solvents, F3O3 may suffer from more severe solvent decomposition, contributing to the more irreversible capacity loss and corresponding deleterious cycling performance. Those observations are consistent with simulation results that the fluorinated ethers are more likely to participate in SEI formation, leading to F-rich components. However, an excess amount of solvent decomposition may not always benefit the SEI stability as well as cycling performance.

In this work, based on a new family of asymmetric HFE solvents, an *in silico* screening protocol was established to accelerate the discovery of HFE solvents through the understanding of the solvation-property relationship, predicting dynamics, redox properties, and possible roles in SEI formations. Analysis of simulation results reveals that the solvation structures of these HFE solvents are highly dependent on the choices of building segments, such as polyethers and fluorocarbons. For instance, HFEs with longer polyether chains are more likely to appear in the first solvation shell around Li ion with decreased number of coordinated TFSI ions and suppress the formation of ionic aggregates. Although the effect of the fluorocarbons on the solution structure is not as significant, increasing the fluorocarbons does impair the solvation effects of solvent molecules and benefits the formation of ionic aggregates. The selected F2O3 electrolyte exhibits a unique solvation structure featuring small dispersive aggregates composed of polyethers in the interstitial space of a percolated network composed of fluorocarbon segments.

By analyzing the solution structures, we then predicted the dynamic and redox properties of the HFE electrolytes. Consistent with solvation structure analysis, electrolytes with longer polyether chains are associated with higher ionic conductivities and shorter lifetimes of ion pairs due to the higher fraction of SSIP and weaker binding between ions, respectively. On the other hand, a higher fraction of CIP and AGG is related to the introduced fluorocarbons, which may benefit the formation of the anion-derived SEI layer. Furthermore, the interfacial reaction during the formation of the SEI layer is elucidated by the AIMD simulation and XPS experiments. We found the decomposition of HFEs is much faster than that of DME, which contributes to the formation of fluorine-based SEI layers. Those results were verified via experimental evaluations of those HFE electrolytes. The electrolyte containing F2O3 indeed shows the highest conductivity, excellent oxidative stability, high ionic conductivity, and superb cycling performance in Li/NMC622 cells.

By adopting an *in-silico* screening protocol, we demonstrated a simulation-guided development cycle for new HFE solvents through the understanding of the solvation-property relationship, predicting dynamic, redox properties, and possible roles in SEI formations. As predicted, the F2O3 solvent demonstrated much-improved dynamics, including high solubility of LiTFSI, much-enhanced conductivity, as well as excellent redox properties, such as improved oxidative stability, benefiting the SEI via solvent participation, collectively contributing to the superb cycling performance in Li/NMC622 cells. This work exemplified a prolific approach of using simulations to correlate fundamental solvation events and desired properties to the chemical structures constructed from building block functional moieties to guide the discovery and rational design of new solvent materials.

ACKNOWLEDGMENTS

This research was supported by the Joint Center for Energy Storage Research (JCESR), a U.S. Department of Energy, Energy Innovation Hub. The submitted manuscript has been created by UChicago Argonne, LLC, Operator of Argonne National Laboratory ("Argonne"). Argonne, a U.S. Department of Energy Office of Science laboratory, is operated under contract no. DE-AC02-06CH11357. We gratefully acknowledge the computing resources provided on Bebop, a high-performance computing cluster operated by the Laboratory Computing Resource Center at Argonne National Laboratory. The authors thank Dr. Kang Xu, Dr. Marshall Schroeder and Dr. Janet Ho from Army Research Laboratory for inspiring discussions regarding the characterization of electrolyte behavior on lithium metal surfaces. T. Li is thankful for the assistance from Bowen An at Northern Illinois University.

Notes

The authors declare no competing financial interest.

# Supporting information

# Simulation Guided Molecular Design of Hydrofluoroether Solvent for High Energy Batteries

Zhou Yu[1,2,+], Zhangxing Shi[1,3,+], Sambasiva R. Bheemireddy[1,3], Ethan Kamphause[1,2], Bowen An[4], Mohammad Afsar Uddin[1,5], Zhiguang Li[1,3,7], Zhenzhen Yang[1,3], Tao Li[1,4,6], Jeffrey S. Moore[1,5,8,9], Lu Zhang[1,3*], Lei Cheng[1,2*]

[1]Joint Center for Energy Storage Research, Argonne National Laboratory, Lemont, IL 60439, USA.
[2]Materials Science Division, Argonne National Laboratory, Lemont, IL 60439, USA
[3]Chemical Sciences and Engineering Division, Argonne National Laboratory, Lemont, IL 60439, USA
[4]Department of Chemistry and Biochemistry, Northern Illinois University, DeKalb, IL 60439, USA
[5]Department of Chemistry, University of Illinois at Urbana–Champaign, Urbana, IL 61801, USA
[6]X-ray Science Division, Argonne National Laboratory, Lemont, IL 60439, USA
[7]Indiana University-Purdue University Indianapolis, 723 West Michigan Street, Indianapolis, IN, 46202, USA
[8]Beckman Institute for Advanced Science and Technology, University of Illinois at Urbana–Champaign, Urbana, IL 61801, USA
[9]Department of Materials Science and Engineering, University of Illinois at Urbana–Champaign, Urbana, IL 61801, USA

[+]Zhou Yu and Zhangxing Shi contributed equally to this work.
*Corresponding Authors: Lu Zhang luzhang@anl.gov and Lei Cheng leicheng@anl.gov

## 1. Theoretical Calculations

Table S1. Classical MD system setups

| Name | Solvent No. | Ion No. | Box (nm) | C (M) |
|---|---|---|---|---|
| F2O2 | 500 | 100 | 5.475 | 1.012 |
| F2O3 | 500 | 120 | 5.838 | 1.001 |
| F3O2 | 500 | 120 | 5.748 | 1.049 |
| F3O3 | 500 | 140 | 6.087 | 1.031 |
| DME | 500 | 60 | 4.659 | 0.985 |

*Classical MD Simulation:* The classical MD simulation system was composed of ~1 M lithium bis(trifluoromethanesulfonyl)imide (LiTFSI) and different solvents, including four newly synthesized fluorinated ethers (e.g., F2O2, F2O3, F3O2, F3O3) and dimethoxyethane (DME) shown in Table S1. The molecular geometries of these fluorinated ethers were generated from scratch and optimized with the universal force field[1] using the Avogadro program.[2] The initial configurations in classical MD simulations were packed using the Packmol code.[3] The simulation box is periodic in all three directions. The force fields of the system were built based on the OPLS-aa force field.[4] Specifically, the force fields of TFSI ion were extracted from the prior work,[5] and the force fields of the solvents were generated using the LigParGen code.[6-7]

The classical MD simulations were performed using the GROMACS code[8] with a 2 fs time step. First, a 50 ns equilibrium simulation was performed under the NPT ensemble for each system. The system energy, box size, and the solvation environment around Li ions are stable during the last 10 ns of the simulation. Then, a 250 ns production run was performed under the NVT ensemble. The Parrinello-Rahman barostat was used to maintain the system pressure at 1 atm with a time constant of 10 ps.[9] The V-rescale thermostat was used to stabilize the system temperature at 350 K with a time constant of 1 ps.[10] The electrostatic interactions were computed using the particle mesh Ewald (PME) method.[11] The real space cutoff and fast Fourier transform spacing were set to 1.2 and 0.12 nm, respectively. The non-electrostatic interactions were computed by direct summation with a cutoff length of 1.2 nm. The last 200 ns trajectory in the production run was used for analysis.

Table S2. AIMD system setups

| Name | Position of salt | DME No. | FXO3 No. | LiTFSI No. | Box (Å$^3$) |
|---|---|---|---|---|---|
| DME_middle | middle | 17 | / | 2 | 14.04×14.04×27.63 |
| DME_interface | interface | 17 | / | 2 | 14.04×14.04×27.63 |
| F2O3_middle | middle | / | 8 | 2 | 14.04×14.04×27.30 |
| F2O3_interface | interface | / | 8 | 2 | 14.04×14.04×27.30 |
| F3O3 | middle & interface | / | 7 | 2 | 14.04×14.04×26.72 |

*Ab Initio MD Simulation:* *Ab initio* MD system is composed of a Li metal electrode and a slab of electrolyte. The most stable (1 0 0) surface has been used for the Li electrode.[12] Meanwhile, the Li electrode has seven layers, and the center three layers have been fixed during the simulation. The simulation box is periodic in all three directions. The space between electrodes is thicker than 16.50 Å. The concentration of LiTFSI in all simulations is ~1 M. The *ab initio* MD system setups are summarized in Table S2. The initial configurations for the *Ab initio* MD simulations with LiTFSI in the middle of the electrolyte or at the interface between electrode and electrolyte were packed using the Packmol code.[3] The system was firstly relaxed through classical MD simulation. The *ab initio* MD simulations were then carried out using the VASP code.[13-14] The projector augmented waves (PAW) method was used to compute the interatomic forces with the Perdew–Burke–Ernzerhof (PBE) generalized gradient approximation for the exchange-correlation energy.[15-16] The plane waves energy cutoff was set as 400 eV. The Brillouin zone was sampled at the Γ-point.[17] The convergence criteria for the electronic self-consistency and ionic relaxation loop were set to 10$^{-5}$ and 10$^{-4}$ eV, respectively. The time step is 1 fs and the temperature is 350 K. A 10 ps trajectory was produced for each system under the NVT ensemble with Nose-Hoover thermostat for the analysis of the interfacial reaction. Meanwhile, we performed Bader charge analysis on each system in order to elucidate the electronic distribution.[18-21] Specifically, 100 frames were extracted from a 10 ps trajectory of each simulation with a time interval of 100 fs, and single point calculations were performed for each frame.

*Quantum Chemical Cluster Calculation:* The molecular geometries were optimized at the DFT level with the B3LYP/6-31+G(d,p) and M052X/6-31+G(d,p) method using the Gaussian program.[22] The energies of the highest occupied molecular orbital (HOMO) and lowest unoccupied molecular orbital (LUMO) were obtained from the optimized geometry.

*Single-linkage clustering algorithm:* For the ionic aggregate analysis, we start from a Li ion and search the O atom in TFSI ions within the cutoff distance, which corresponds to the location of the first valley in the RDF from Li ion to O atom in TFSI ion shown in Figure S2. If no O atoms were found, we choose another Li ion and carry out the previous step. Otherwise, we consider all four O atoms in the newly found TFSI ion as the center atoms and search for the other neighboring Li ions within the cutoff distance. The process is repeated until no new Li ion is found.

For the aggregates composed of F/O in solvent molecules, we start from a F/O atom in a solvent molecule and search for other F/O atoms in solvent molecules within the cutoff distance, which corresponds to the location of the first valley in the RDF between F/O in solvent molecules shown in Figure S6. Then we regard the newly found F/O atoms as the center atoms and repeat the previous step. If no new F/O atoms were found, the search process was conducted from another F/O atom until there is no new F/O atom in solvent molecules.

*Li-TFSI association correlation function:* To characterize the lifetime of Li-TFSI ion pair in different electrolytes (see Figure S9), we calculated the Li-TFSI association correlation function following $\text{ACF}(t) = \langle c(0)c(t) \rangle$, where $c(t)$ is an indicator of the Li-TFSI association. $c(t)$ is 1.0 if a TFSI ion coordinating with a Li ion at time 0 is continuously stay in the first solvation shell of this Li ion by time t. The correlated function can be fitted to the stretched exponential function following $\text{ACF}(t) = a_0 \exp(-(t/a_1)^{a_2})$. Then, the lifetime of the Li-TFSI ion pair is calculated according to the fitting parameters $a_0$, $a_1$, and $a_2$ following the gamma function $\tau_{ij} = a_0 a_1 \Gamma(1 + 1/a_2)$.

*Diffusivity and conductivity calculation:* The self-diffusivity of ions and solvents and the ionic conductivity of electrolytes are calculated according to the following equations: $D_A = \frac{1}{6t} \lim_{t \to \infty} \langle |R_i(t) - R_i(0)|^2 \rangle_{i \in A}$ and $\sigma = \lim_{t \to \infty} \frac{e^2}{6tVk_BT} \sum_{ij}^{N} Z_i Z_j \langle [R_i(t) - R_i(0)][R_j(t) - R_j(0)] \rangle$. Where $D_A$ is the self-diffusivity of A, $R_i(t)$ is coordinates of $i$ at time t, $\sigma$ is the ionic conductivity, $V$ is the volume of the system, $T$ is temperature, and $Z_i$ is the charge on $i$.

## 2. Materials and Syntheses

Fluorinated ethers were obtained by using the following general synthesis procedure. To a suspension of NaH (1 eq) in DMF was added fluorinated alcohol (1 eq) dropwise under a nitrogen atmosphere at 0 °C. This mixture was stirred at 0 °C for 20 minutes and at room temperature for 10 minutes. Then, Bromo ether (0.5 eq) starting material was added to the reaction mixture dropwise at room temperature, and the reaction was stirred overnight before water was added. The final products were obtained by diethyl ether extraction.

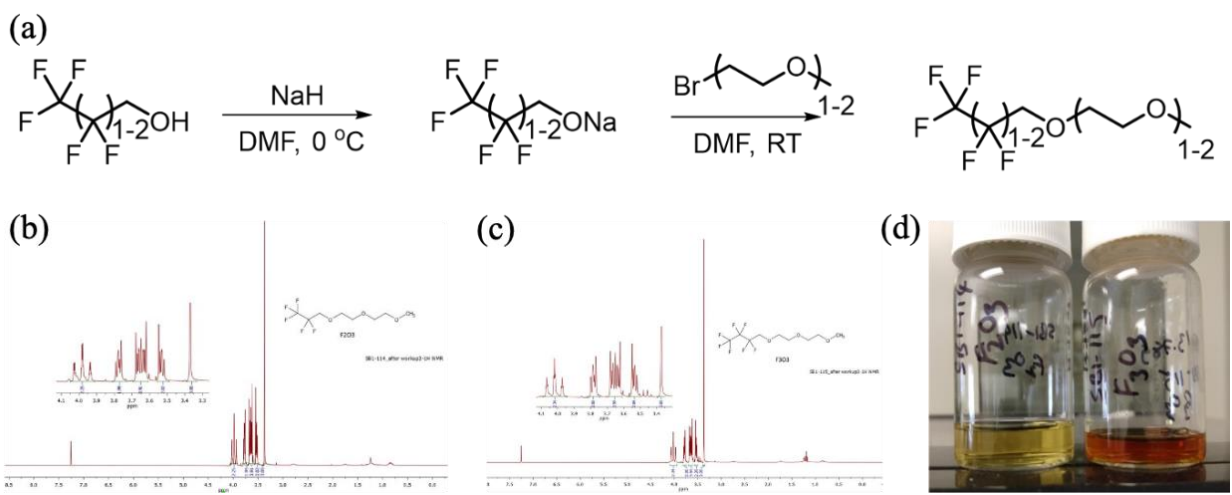

Figure S1. (a) Representative synthetic scheme.

## 3. Material Characterizations

*Raman spectroscopy.* Raman experiments were conducted with a Renishaw confocal micro-Raman system at the Center for Nanoscale Materials of Argonne National Laboratory. The samples were first sealed in quartz capillaries and measured within the capillaries. The quartz capillary has no Raman bands, so background correction is not required. Raman excitation was applied with laser wavelengths of 532 nm. Raman spectra were recorded using a 50× focusing/collection optic with a numerical aperture of 0.5 (Leica).

*X-ray photoelectron spectroscopy.* After rinsing the electrodes, the samples were transferred, without air exposure, to the XPS chamber (Physical Electronics), which is attached to an Ar-atmosphere glovebox. The high resolution spectra were obtained using a 100 μm beam (25 W) with Al Kα radiation (hυ = 1486.6 eV), $Ar^{+-}$ ion and electron beam sample neutralization, fixed analyzer transmission mode, and pass energy of 23.25 eV. For data analysis of all spectra, the Shirley background was subtracted, and the resulting spectrum was fitted to multiple Gaussian peaks using Multipak software provided by Physical Electronics.

## 4. Electrochemical Measurements

The cycling performance of lithium metal cells was evaluated using 2032-type stainless steel coin cells. The cells were configured with a lithium metal electrode, a microporous polypropylene separator (Celgard 2325), an NMC 622 cathode, and 25 μL of electrolyte containing 10 wt% FEC. The cells were first subjected to three formation cycles at a C/10 rate followed by 100 cycles at a C/3 rate with the cell voltage maintained between 3.0 to 4.2 V during this cycling. The current rate was derived from the actual capacity during the formation cycle.

The ionic conductivity of electrolytes was measured by conducting electrochemical impedance spectroscopy (EIS, Figure S14) on an assembled testing cell. A 2032-coin cell was assembled consisting of a Teflon ring sandwiched between two stainless steel spacers. Electrolyte samples were filled in the space between the spacers. The EIS was con in the frequency range from 1 MHz to 0.1 Hz. Ionic conductivity can be calculated using the Z' intercept at a high frequency range obtained from the Nyquist plot of the EIS spectrum. F2O3 and F3O3 electrolytes show ionic conductivity values of 1.81 and 0.76 mS/cm, respectively.

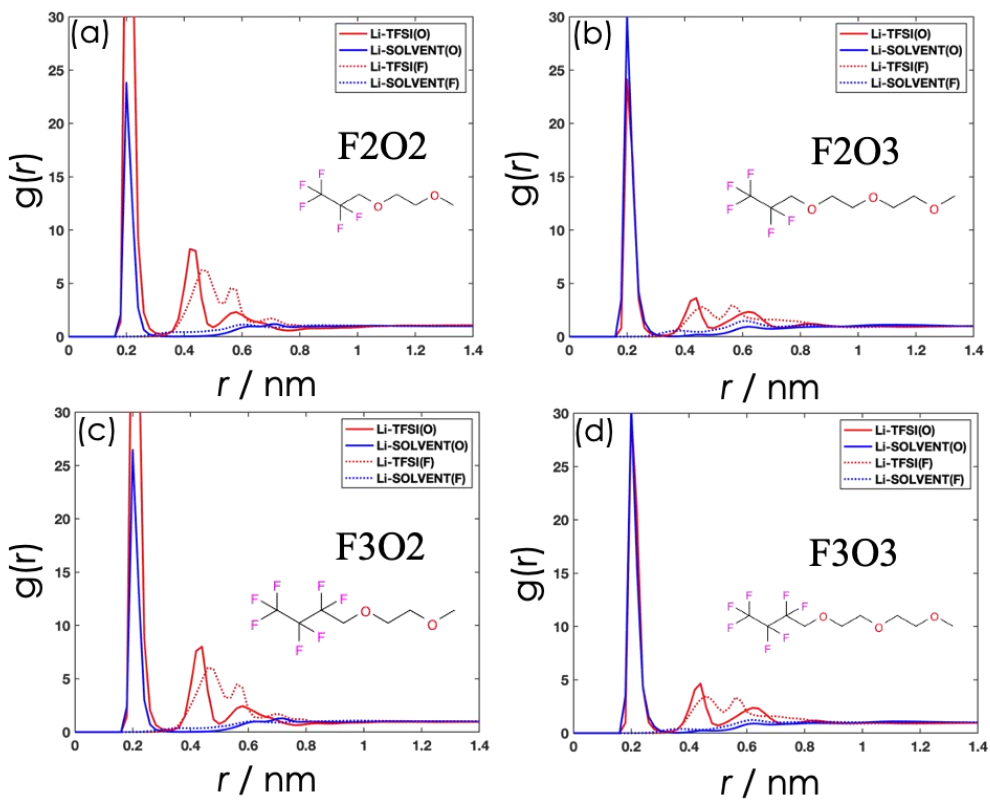

Figure S2. Radial distribution function (RDF) from Li ion to O and F in TFSI ion and fluorinated ethers.

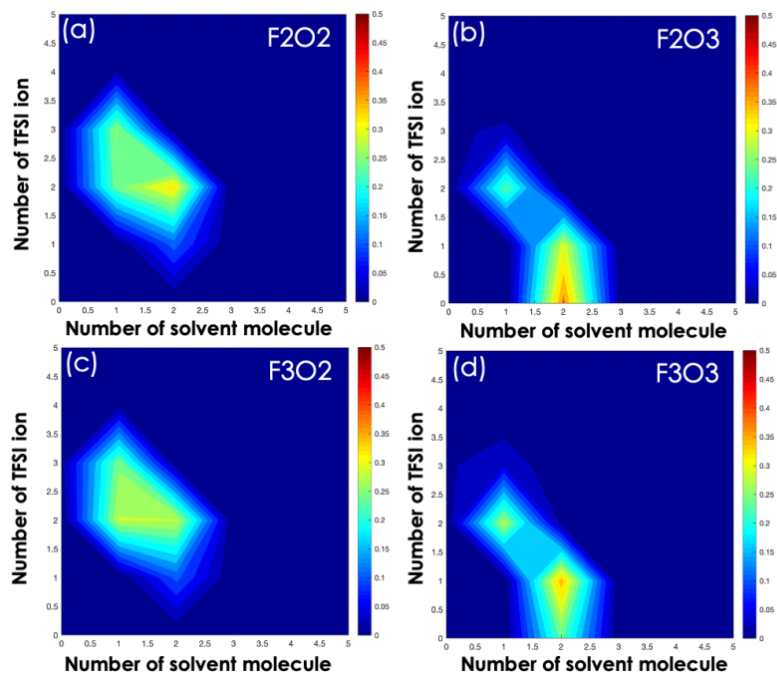

Figure S3. Time-averaged appearance frequency of the coordination number (CN) of TFSI ions and solvent molecules in the first solvation shell of Li ion in different electrolytes. The color bar represents the occurrence frequency.

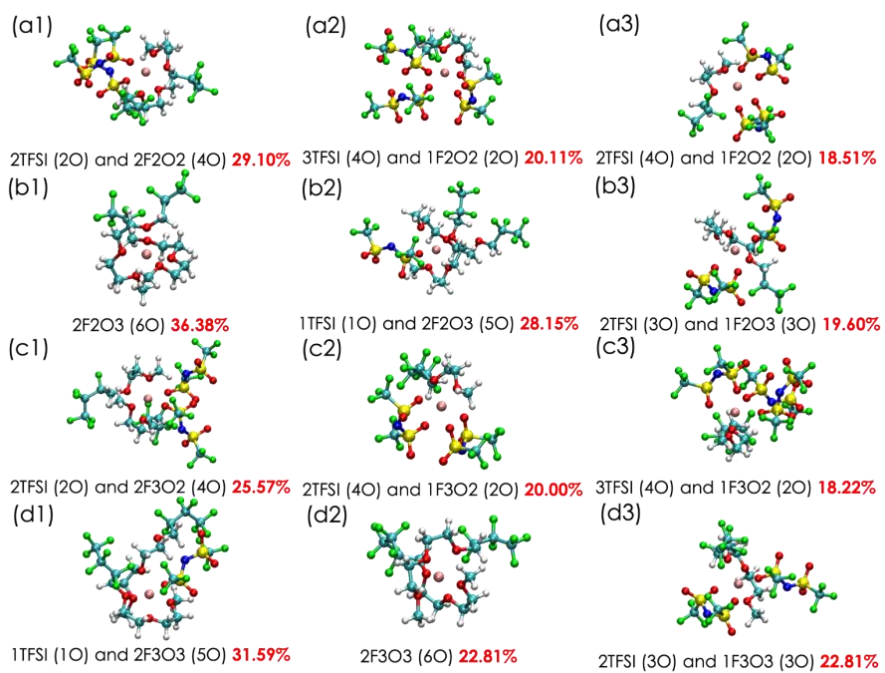

Figure S4. Representative solvation structures around Lis in F2O2, F2O3, F3O2, and F3O3 systems, respectively. The fraction of each solvation structures is marked in red. Pink, blue, yellow, red, cyan, and green balls denote Li, N, S, O, C, and F, respectively.

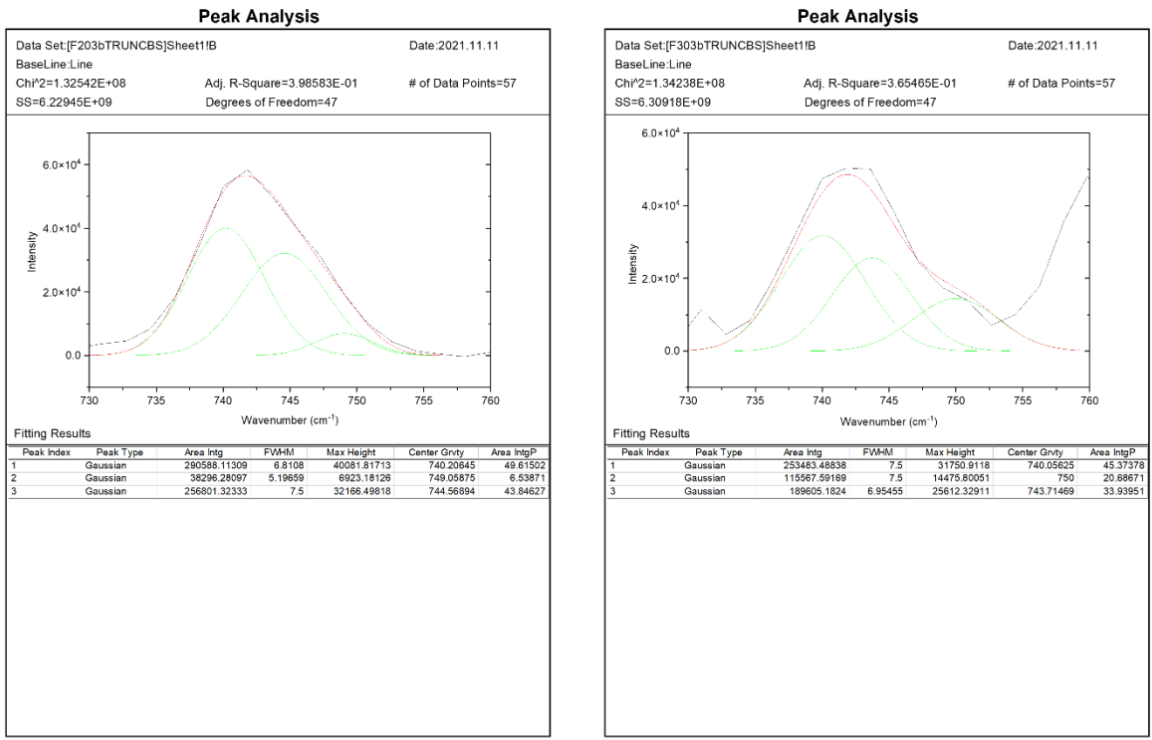

Figure S5. Raman Spectral deconvolution for 1 M LiTFSI in F2O3 and F3O3.

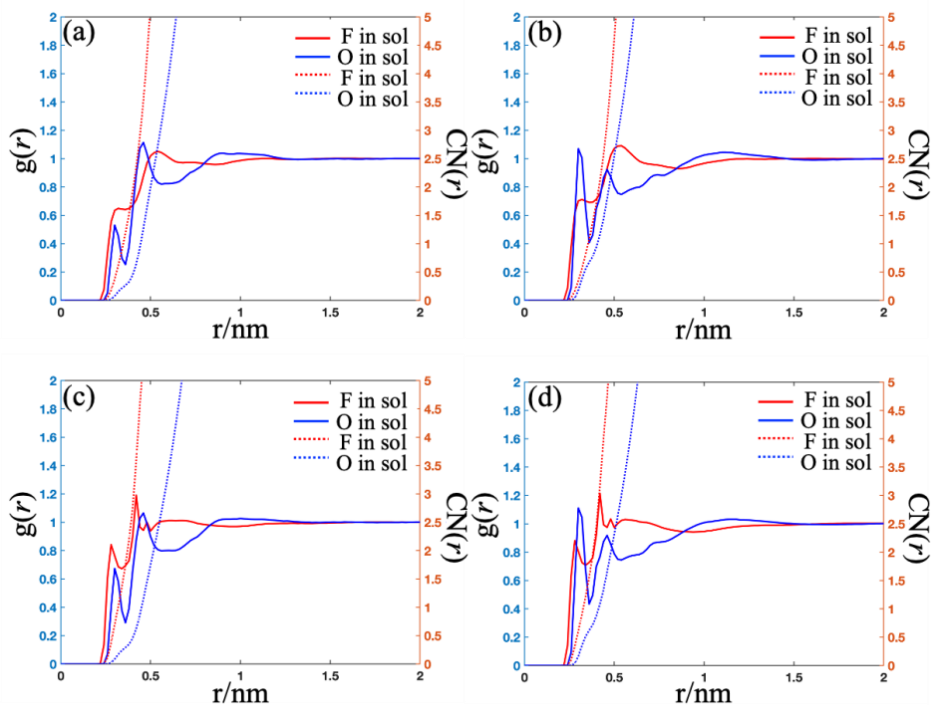

Figure S6. RDF and CN between F or O in the electrolytes with (a) F2O2, (b) F2O3, (c) F3O2, and (d) F3O3. The intramolecular F or O interaction is excluded.

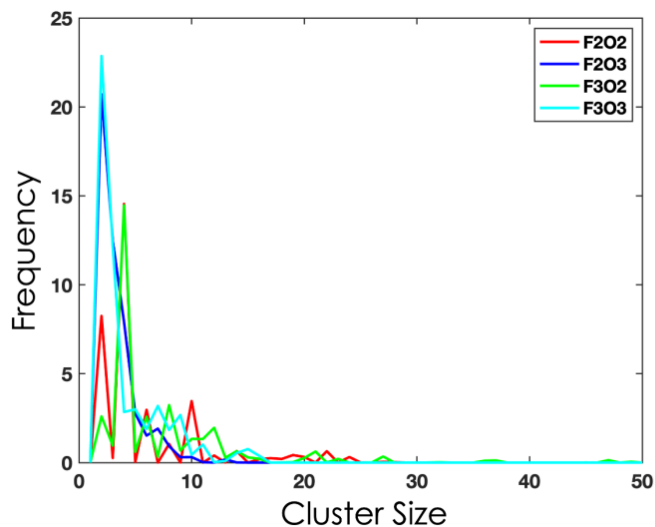

Figure S7 Time-averaged appearance frequency of ionic aggregates composition in different electrolytes. 500 frames were extracted from a 20 ns trajectory of each simulation with a time interval of 40 ps.

Table S3. Electrolytes nanostructure

|      | MaxF/TotF Cutoff=0.36 nm | AveO Cutoff=0.36 nm | AveO Cutoff=0.46 nm | AveIon | Free solvent |
|------|--------------------------|---------------------|---------------------|--------|--------------|
| F2O2 | 95.77%                   | 2.28                | 8.00                | 5.88   | 70.93%       |
| F2O3 | 85.54%                   | 3.58                | 34.06               | 1.69   | 59.66%       |
| F3O2 | 99.37%                   | 2.30                | 5.77                | 7.18   | 66.79%       |
| F3O3 | 97.90%                   | 3.60                | 21.19               | 2.28   | 55.74%       |

*MaxF/TotF represents the fraction of F atoms in solvent molecules forming the maximal cluster. AveO and AveIon represents the average number of O atoms in solvent molecules and ions (e.g., $Li^+$ and $TFSI^-$ ion) in each cluster, respectively.

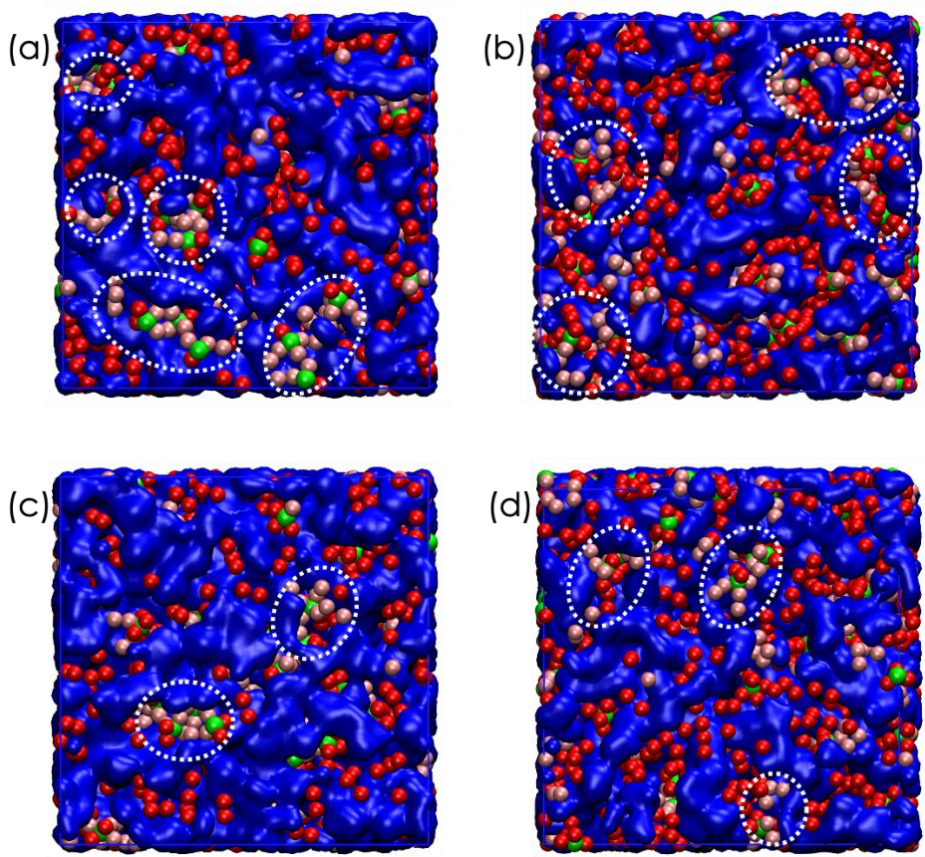

Figure S8. Snapshot of solution structure in 1 M LiTFSI (a) F2O2 (b) F2O3 (c) F3O2 (d) F3O3 electrolytes. Blue region represents F in solvent molecules and TFSI ions. Green, red, and pink balls denote Li, O in solvent and TFSI ion, respectively. White dotted cycles indicate some localized high concentration regions.

Table S4. Transport properties in different electrolytes. L represents the characteristic length.

| Name | D-Li ($10^{-11}$ m$^2$/s) | D-TFSI ($10^{-11}$ m$^2$/s) | D-Solvent ($10^{-11}$ m$^2$/s) | Conductivity (mS/cm) | Lifetime (ns) | L (Å) |
|---|---|---|---|---|---|---|
| F2O2 | 4.77±0.07 | 4.53±0.05 | 38.94±0.10 | 0.1950±0.0114 | 28.69 | 28.65 |
| F2O3 | 3.07±0.02 | 3.21±0.01 | 9.41±0.12 | 1.1177±0.0023 | 3.76 | 8.33 |
| F3O2 | 3.06±0.06 | 2.94±0.04 | 20.54±0.07 | 0.1392±0.0080 | 29.47 | 23.26 |
| F3O3 | 0.94±0.01 | 1.16±0.01 | 4.31±0.02 | 0.3873±0.0010 | 7.79 | 6.63 |

Vehicular and structural motions are two representative transport mechanisms of ions in electrolytes. Vehicular motion represents cation can diffuse with its solvation shell as one species. Structural motion means the cation can diffuse through the continuous exchange of its solvation shell composed of anion and solvent molecules. The transport mechanism of a Li ion as the reference of the TFSI ion in its solvation shell is determined by the characteristic length following $L_{\text{Li-TFSI}} = \sqrt{6D_{\text{Li}}\tau_{\text{Li-TFSI}}}$, where $D_{\text{Li}}$ is the self-diffusivity of Li ion and $\tau_{\text{Li-TFSI}}$ is the lifetime of the Li-TFSI ion pair. Essentially, $L_{\text{Li-TFSI}}$ describe how long the Li-TFSI ion pair can move before the dissociation of the ion pair occur. Therefore, smaller $L_{\text{Li-TFSI}}$ means the diffusion of Li ion tends to apply the structure motion through the continuous exchange of its associated TFSI ion. From Table S4, we can see the increase of O-segments length makes the diffusion of Li ion shift to the structural motion. This is because Li ion is coordinated by more O from solvent and less O from TFSI with the increase of solvent's O-segment length. The binding between Li and TFSI ion weakens, which helps the dissociation of the ion pair.

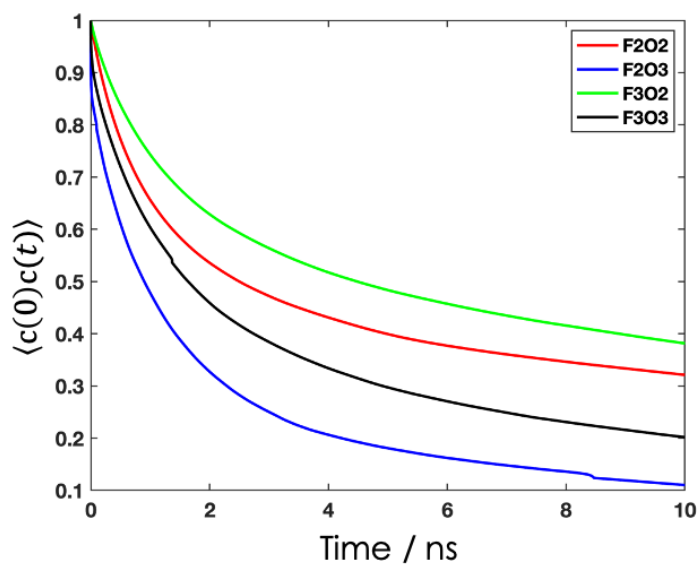

Figure S9. Li-TFSI association correlation function in electrolytes with different fluorinated ethers.

Table S5. LUMO and HOMO energy of the fluorinated ethers and DME

|      | B3LYP/6-31+G(d,p) | | M052X/6-31+G(d,p) | |
|------|-----------|-----------|-----------|-----------|
|      | LUMO (eV) | HOMO (eV) | LUMO (eV) | HOMO (eV) |
| F2O2 | -0.275 | -7.488 | -0.346 | -9.414 |
| F2O3 | -0.266 | -7.244 | -0.337 | -9.167 |
| F3O2 | -0.275 | -7.493 | -0.346 | -9.421 |
| F3O3 | -0.270 | -7.338 | -0.341 | -9.292 |
| DME  | 0.350  | -7.136 | 1.551  | -9.059 |

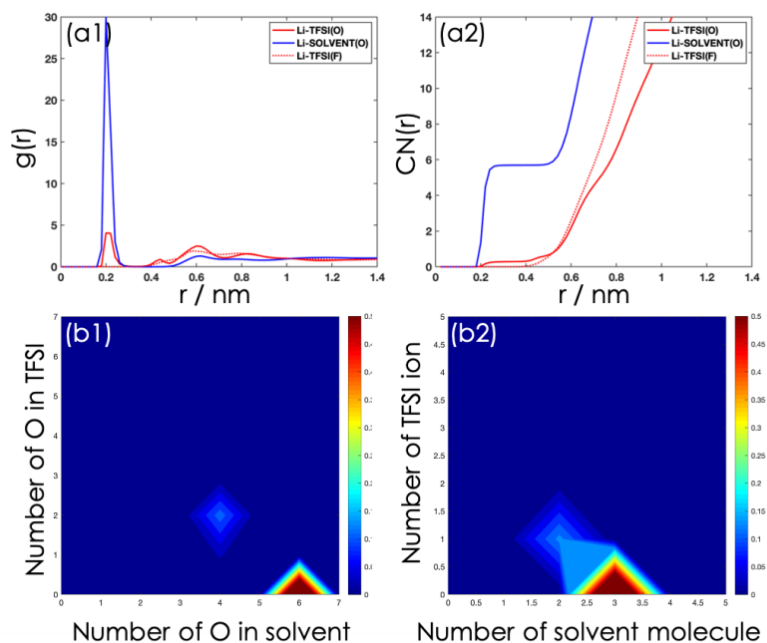

Figure S10. (a1) Radial distribution function from Li ion to O in TFSI and DME and F in TFSI. (a2) Coordination number of O in TFSI and DME and F in TFSI around Li ion. Time-averaged appearance frequency of the CN of (b1) O in TFSI ions and solvent molecules or (b2) TFSI ions and solvent molecules around Li ion in 1 M LiTFSI DME electrolyte. The fraction of free/SSIP, CIP, and AGG in the DME electrolyte is 84.18%, 12.45%, and 3.37%, respectively.

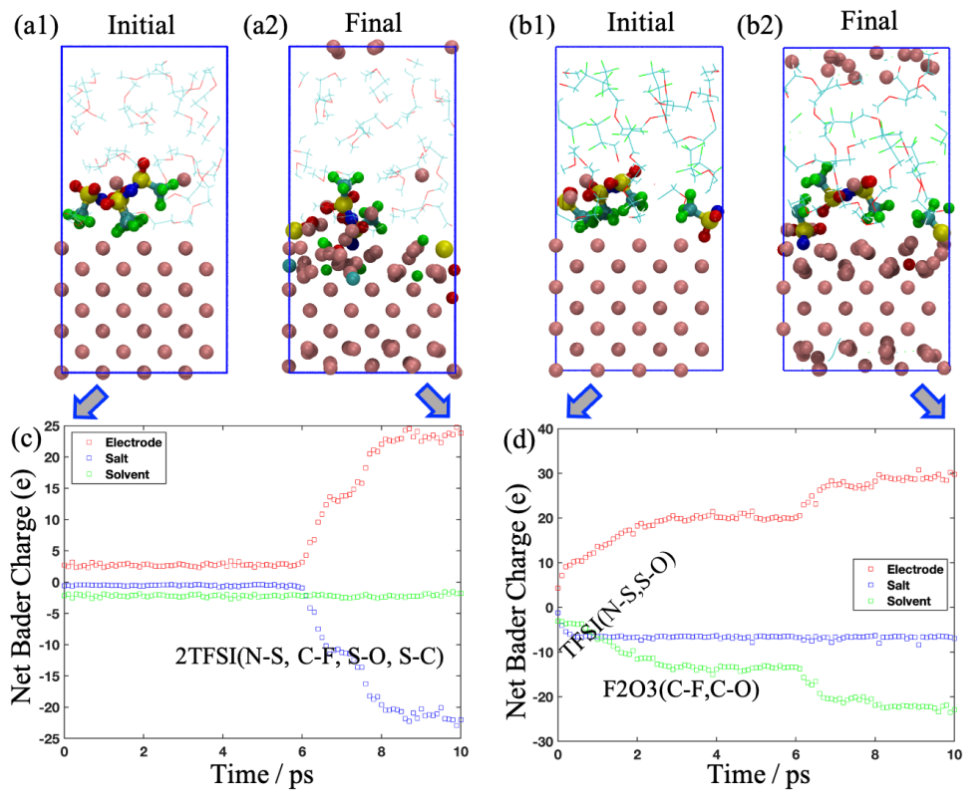

Figure S11. The initial and final snapshots extracted from AIMD simulations with (a) DME as solvent or (b) F2O3 as solvent. LiTFSI locates at the interface between electrode and electrolyte in the initial configuration. Pink, blue, yellow, red, cyan, and green balls denote Li, N, S, O, C, and F, respectively. The solvent molecules are depicted as wireframe. (c-d) Net Bader charges of system components.

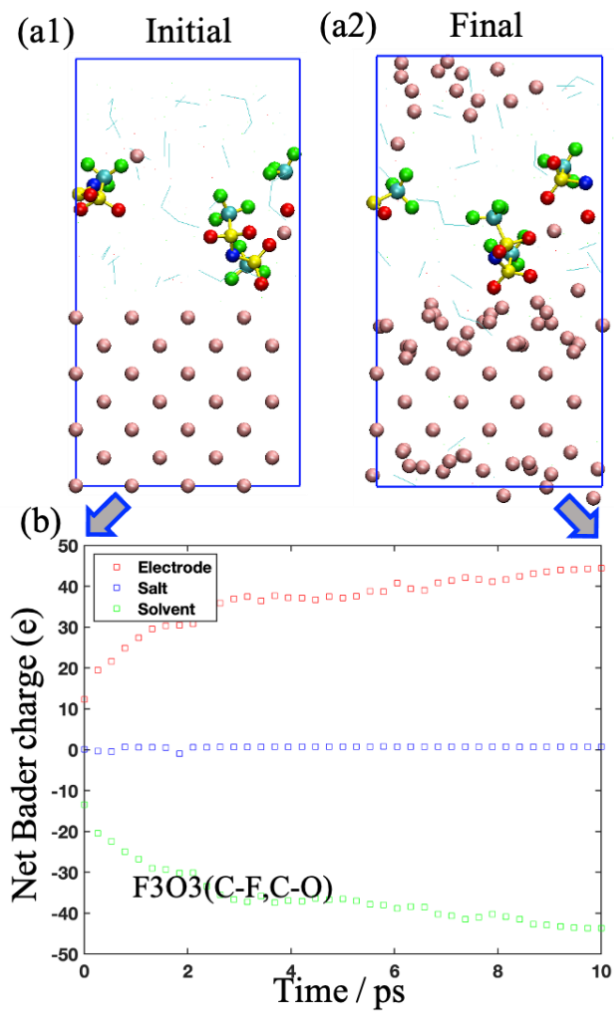

Figure S12. The initial and final snapshots extracted from AIMD simulations with F3O3 as solvent. Pink, blue, yellow, red, cyan, and green balls denote Li, N, S, O, C, and F, respectively. The solvent molecules are depicted as wireframe. (b) Net Bader charges of system components.

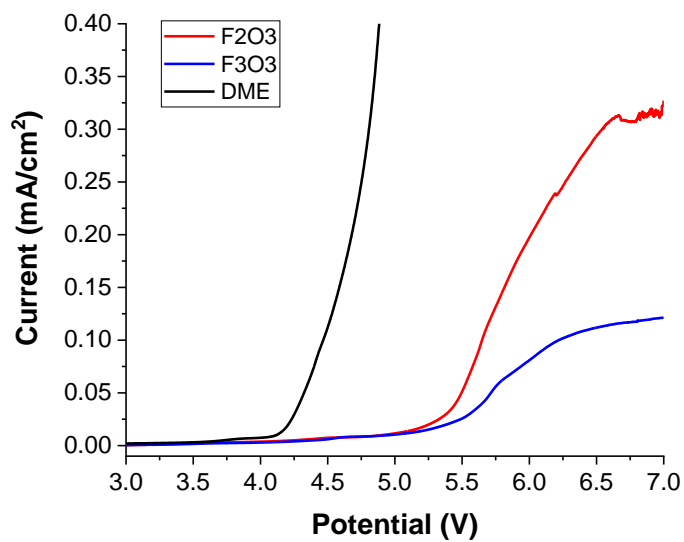

Figure S13. Linear sweep voltammetry (LSV) tests in 2032-type coin cell consists of a Li metal anode, a Celgard 2325 separator, a Al foil as counter electrode and 1 M LiTFSI in F2O3, F3O3 or DME electrolyte. The cells were subjected to a constant voltage sweep from 2.7 V to 7.0 V at a rate of 10 mV/s. F2O3 electrolyte shows excellent oxidative stability with no obvious decomposition until 5.5 V.

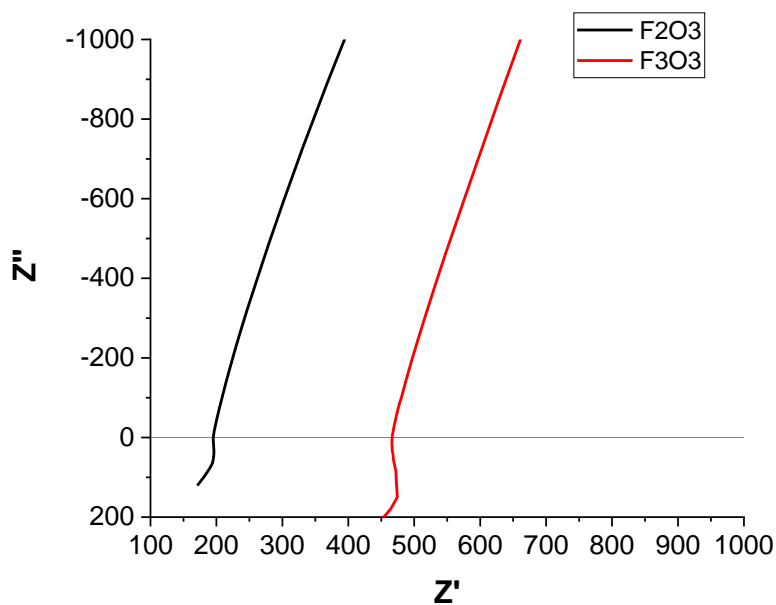

Figure S14. The ionic conductivity of the F2O3 and F3O3 electrolyte was measured using electrochemical impedance spectroscopy (EIS). The 2032-type coin cell used for the test consist of a Teflon ring that filled with F2O3 electrolyte that is sandwiched between two stainless steel current collectors. The electrochemical impedance of the cell was tested in the frequency range from 1 MHz to 0.1 Hz. Ionic conductivity can be calculated using the Z' intercept at a high frequency range obtained from the Nyquist plot of the EIS spectrum. F2O3 electrolyte shows high ionic conductivity of 1.81 mS/cm.

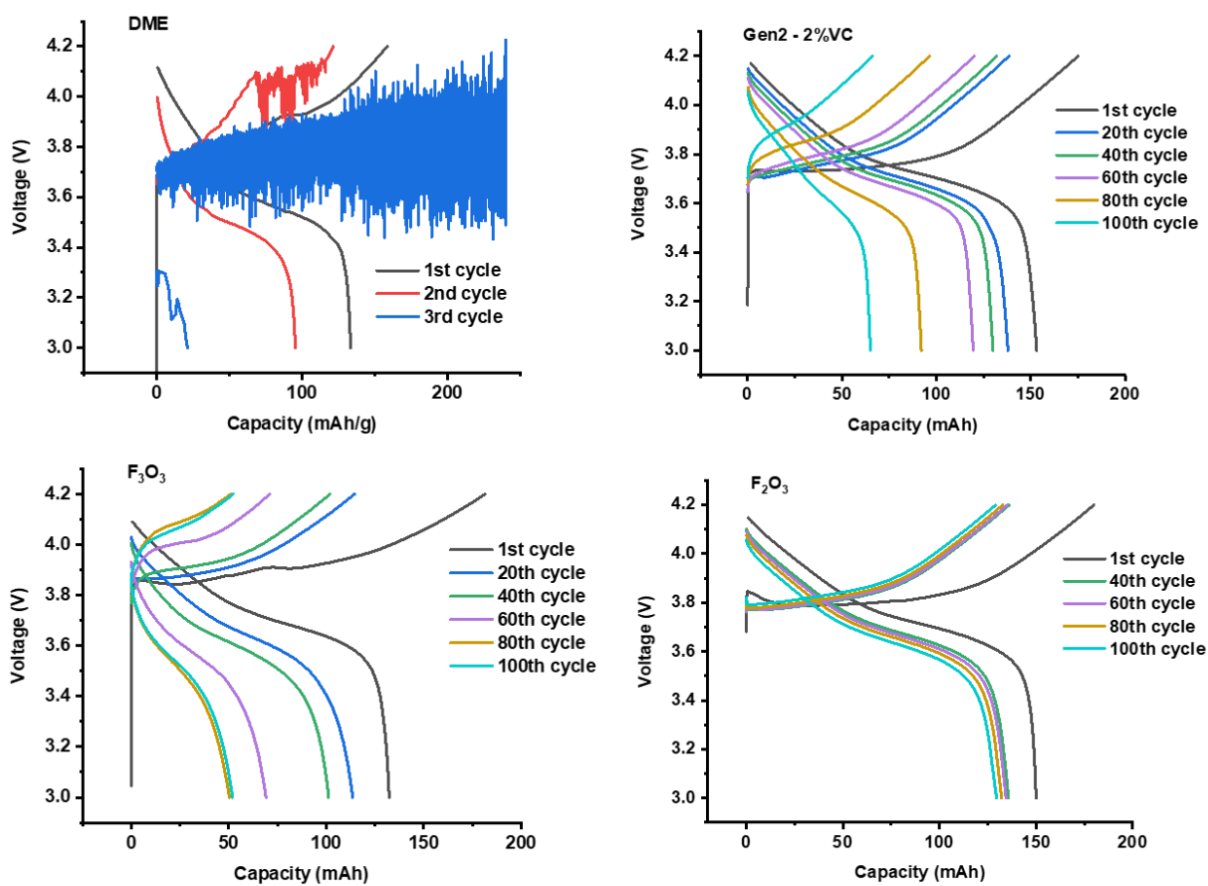

Figure S15. Voltage-capacity profiles for cells containing 1 M LiTFSI DME/F2O3/F3O3 electrolytes and 1 M LiPF6 in EC/EMC (3/7) plus 2% vinylene carbonate electrolyte.

[21] Sanville, E.; Kenny, S. D.; Smith, R.; Henkelman, G., Improved Grid-Based Algorithm for Bader Charge Allocation. *Journal of computational chemistry* **2007,** *28*, 899-908.

[22] Frisch, M.; Trucks, G.; Schlegel, H.; Scuseria, G.; Robb, M.; Cheeseman, J.; Scalmani, G.; Barone, V.; Petersson, G.; Nakatsuji, H., Gaussian 16, Gaussian. *Inc., Wallingford CT* **2016,** *2016*.
17